\documentclass[fleqn,twoside]{article}
\usepackage[headings]{espcrc2}
\usepackage{axodraw}
\usepackage{amsfonts}

\readRCS
$Id: espcrc2.tex,v 1.2 2004/02/24 11:22:11 spepping Exp $
\usepackage{graphicx}
\newcommand{\la}[1]{\label{#1}}
\newcommand{\be}{\begin{equation}}
\newcommand{\ee}{\end{equation}}
\newcommand{\ba}{\begin{eqnarray}}
\newcommand{\ea}{\end{eqnarray}}

\newcommand{\rmii}[1]{{\mbox{\tiny\rm{#1}}}}
\newcommand{\fig}{Fig.~}
\newcommand{\eq}{Eq.~}
\newcommand{\se}{Sec.~}

\newcommand{\nr}[1]{(\ref{#1})}

\newcommand{\nn}{\nonumber \\}
\newcommand{\e}{\epsilon}
\renewcommand{\(}{\left(}
\renewcommand{\)}{\right)}
\newcommand{\lk}{\left[}
\newcommand{\rk}{\right]}
\def\sumint{\hbox{$\sum$}\!\!\!\!\!\!\!\int}

\newcommand{\zA}{Z_1}
\newcommand{\zC}{Z_3}
\newcommand{\ZZ}{{\mathbb{Z}}}
\newcommand{\RR}{{\mathbb{R}}}
\newcommand{\gE}{\gamma_{\rmii E}}
\newcommand{\fra}[2]{\mbox{\small$\frac{#1}{#2}$}}

\newcommand{\pic}[1]{\;\parbox[c]{30pt}{\begin{picture}(30,30)(0,0)
\SetWidth{1.0}\SetScale{1.0} #1 \end{picture}}\;}
\newcommand{\picb}[1]{\;\parbox[c]{45pt}{\begin{picture}(45,30)(0,0)
\SetWidth{1.0}\SetScale{1.0} #1 \end{picture}}\;}
\def\ToptVS(#1,#2,#3){\pic{#1(15,15)(15,0,180) #2(15,15)(15,180,360)%
 #3(30,15)(0,15)}}
\def\ToprVB(#1,#2,#3,#4){\picb{#1(30,15)(15,-120,120) #2(30,15)(15,120,240)%
 #3(15,15)(15,60,300) #4(15,15)(15,-60,60)}}
\def\ToprVBdot(#1,#2,#3,#4){\picb{#1(30,15)(15,-120,120) #2(30,15)(15,120,240)%
 #3(15,15)(15,60,300) #4(15,15)(15,-60,60)\Vertex(0,15){2}}}
\def\ToprVM(#1,#2,#3,#4,#5,#6){\pic{#3(15,15)(15,-30,90) #1(15,15)(15,90,210)%
 #2(15,15)(15,210,330) #5(2,7.5)(15,15) #6(15,15)(15,30) #4(28,7.5)(15,15)}}
\def\ToprVMdot(#1,#2,#3,#4,#5,#6){\pic{#3(15,15)(15,-30,90)%
 #1(15,15)(15,90,210)%
 #2(15,15)(15,210,330) #5(2,7.5)(15,15) #6(15,15)(15,30) #4(28,7.5)(15,15)%
 \Vertex(23,3){2}\Vertex(15,22.5){2}}}
\def\TopfVT(#1,#2,#3,#4,#5,#6){\pic{#1(15,15)(15,90,210)%
 #2(15,15)(15,210,330) #3(15,15)(15,-30,90) #4(2,7.5)(15,30)%
 #6(28,7.5)(2,7.5) #5(15,30)(28,7.5)}}
\def\Lsc(#1,#2)(#3,#4){\Line(#1,#2)(#3,#4)}
\def\Lqu(#1,#2)(#3,#4){\ArrowLine(#1,#2)(#3,#4)}
\def\Luq(#1,#2)(#3,#4){\ArrowLine(#3,#4)(#1,#2)}
\def\Asc(#1,#2)(#3,#4,#5){\CArc(#1,#2)(#3,#4,#5)}
\def\Aqu(#1,#2)(#3,#4,#5){\ArrowArc(#1,#2)(#3,#4,#5)}
\def\Auq(#1,#2)(#3,#4,#5){\ArrowArcn(#1,#2)(#3,#5,#4)}
\def\Ahh(#1,#2)(#3,#4,#5){\DashCArc(#1,#2)(#3,#4,#5){1}}
\def\Lhh(#1,#2)(#3,#4){\DashLine(#1,#2)(#3,#4){1}}
\def\Agh(#1,#2)(#3,#4,#5){\DashArrowArc(#1,#2)(#3,#4,#5){1}}
\def\Lgh(#1,#2)(#3,#4){\DashArrowLine(#1,#2)(#3,#4){1}}

\title{Loops for Hot QCD}

\author{Y. Schr\"oder\address[MCSD]{Theoretical Physics, 
        University of Bielefeld, 
        PO Box 100131, 33501 Bielefeld, Germany}\thanks{Talk given at 
the International Conference on Loops and Legs in Quantum Field Theory 2008}}
       
\begin{document}

\begin{abstract}
In this talk we review the status concerning vacuum integrals 
needed in perturbative expansions of QCD at non-zero temperature.
We will focus on the differences as compared to familiar 
zero-temperature techniques, and provide a list of known basic master 
integrals.
\vspace{1pc}
\end{abstract}

\maketitle

\section{Introduction}

Weak-coupling expansions in finite temperature field theory can be organized
in a way that is very similar to the better-known zero-temperature case. 
In particular, for
observables describing the physics of thermally equilibrated systems, the
so-called imaginary time formalism provides a very close analogy. 

For a given system that is coupled to an external heat-bath with temperature
$T$, the relevant observables have to be averaged with the density matrix,
$\langle {\cal O} \rangle = 
{\rm Tr}\({\cal O}\,e^{-H/T}\)/{\rm Tr}\(e^{-H/T}\)$, 
where $H$ denotes the Hamiltonian of the system. 
In the case of interacting quantum fields $\phi$ in thermal equilibrium, 
one can lift this to a functional integral representation of the averages, 
\ba
\langle {\cal O}[\phi] \rangle &=& 
\frac{\int{\cal D}\phi\,{\cal O}[\phi]\,e^{-\int_0^{1/T}d\tau 
\int d^3x {\cal L}_{\rmii E}[\phi]}}
{\int{\cal D}\phi\,e^{-\int_0^{1/T}d\tau 
\int d^3x {\cal L}_{\rmii E}[\phi]}}\;,
\ea
with Euclidean action and
where the trace is enforced by constraining the path integral to fields 
that are periodic in {\em imaginary time} $\tau$: 
\ba
\phi(\tau+1/T,\vec x)=\pm\phi(\tau,\vec x)\;,
\ea 
where the upper/lower sign
refers to bosonic/fermionic fields.

Finite-temperature Feynman rules hence differ from zero-temperature ones 
by the fact that, due to the compact support in imaginary time $\tau$,
the integration measure includes a discrete sum 
$\int \frac{dp_0}{2\pi}\rightarrow T\sum_{p_0}$ over the
{\em Matsubara frequencies} $p_0=2\pi T n$ for bosons and
$p_0=2\pi T (n+\frac12)$ for fermions, respectively, where $n\in\ZZ$.
As the presence of a preferred direction -- the rest frame of the heat-bath -- 
has broken the 4d Lorentz symmetry, one works with four-momenta 
$P=(p_0,\vec p)$ and uses Euclidean metric $g_{\mu\nu}=\delta_{\mu\nu}$, such 
that $P^2=p_0^2+\vec p^2$.

In this short note, we will assume the reader be familiar with standard 
methods of zero-temperature perturbation theory, 
such as integration by parts (IBP) \cite{ibp} and its systematic 
algorithmic use \cite{lapo},
and then point out the main
differences that occur in a thermal system. For illustration purposes,
we will concentrate on the most basic class of sum-integrals, 
vacuum (or bubble) integrals, which are needed to compute e.g. thermodynamic
observables that directly relate to vacuum diagrams, 
like the pressure (see \cite{BN,phi4} and references 
therein).  
These vacuum integrals are at the same time needed as building blocks 
for higher-point Greens functions, after expanding in external momenta,
see e.g. \cite{2loopG}.
A collection of analytically known integrals that are relevant in these
contexts will be presented.

\section{Bosonic master integrals}
\la{se:bos}

We work in dimensional regularization,
and for convenience take the integral measure to be
\ba
\sumint_P \equiv T\sum_{n=-\infty}^\infty\,
\(4\pi T^2\)^\e \int \frac{d^{3-2\e}p}{(2\pi)^{3-2\e}} \;.
\ea

\subsection{One-loop bosonic integrals}

The basic bosonic 1-loop tadpole integral is
\ba \label{eq:bosTad}
{\cal I}_n^m &\equiv& \sumint_P \frac{(p_0)^m}{(P^2)^n} \nn
&=& \frac{2\pi^2 T^4}{(2\pi T)^{2n-m}}
\frac{4^\e\,\Gamma(n-\frac32+\e)}{\Gamma(\frac12)\Gamma(n)}\,
\times\nn &\times&
\zeta(2n-m-3+2\e) \;.
\ea
It obeys the recursion relation 
${\cal I}_{n+1}^{m+2}=\frac{2n-3+2\e}{2n}\,{\cal I}_n^m$, as can either
be derived from integration by parts (IBP) in the integral over 3-momentum, 
or confirmed directly from the above explicit solution.
Note that, in contrast to the zero-temperature case, already at 1-loop level
one finds in principle infinitely many master integrals. 
In practice, however, only a small number of them appear in any 
practical computation.

1-loop master integrals that are needed for the pressure are
\ba
{\cal I}_1^0 &\!\!=\!\!& \frac{T^2}{12}\;
\frac{4^\e\,\Gamma(\frac12+\e)}{\Gamma(\frac12)}\,
\frac{\zeta(-1+2\e)}{\zeta(-1)}\,\frac1{1-2\e}\nn
&\!\!\approx\!\!& \frac{T^2}{12}
\lk 1+\e\(2-\gE+2\zA\)+...\rk \nonumber \;,
\ea
\ba
{\cal I}_2^0 &\!\!=\!\!& \frac1{(4\pi)^2}\,\frac1\e\;
\frac{4^\e\,\Gamma(\frac12+\e)}{\Gamma(\frac12)}\,
2\e\,\zeta(1+2\e)\nn
&\!\!\approx\!\!& \frac1{(4\pi)^2}\,\frac1\e\lk 1+\e\,\gE+...\rk \nonumber \;,
\ea
\ba
{\cal I}_3^0 &\!\!=\!\!& \frac{2\zeta(3)}{(4\pi)^4 T^2}\;
\frac{4^\e\,\Gamma(\frac12+\e)}{\Gamma(\frac12)}\,
\frac{\zeta(3+2\e)}{\zeta(3)}\,(1+2\e)\nn
&\!\!\approx\!\!& \frac{2\zeta(3)}{(4\pi)^4 T^2}
\lk 1+\e\( 2-\gE+2\frac{\zeta'(3)}{\zeta(3)}\)+...\rk \nonumber \,,
\ea
where $Z_i\equiv\frac{\zeta'(-i)}{\zeta(-i)}$.

\subsection{Two-loop bosonic integrals}

At 2-loop order, there is a major simplification: the basic sunset-type
master vacuum integral (cf. \fig\ref{fig:bosInts}) vanishes identically, 
\ba
{\cal S} &\equiv& \sumint_{PQ} \frac1{P^2\,Q^2\,(P-Q)^2} 
\;=\; 0 \la{sunset}\;,
\ea
as can be proven by IBP. In fact, using that the integral of a total
derivative vanishes in dimensional regularization,
\ba
0 &=& \sumint_{PQ} \partial_{p_i}\,f_i\,
\frac1{P^2\,Q^2\,(P-Q)^2} \;,
\ea
and choosing 
(here, $d=4-2\e$ denotes the space-time dimension)
\ba
f_i &=& (d-3)(p_i+q_i)+
\nn&&+
\frac2{Q^2}\,(p_0+q_0)(p_0 q_i-q_0 p_i) \;,
\ea
after working out the derivatives, 
using the shift $Q\rightarrow P-Q$, and exploiting
symmetry of the sum-integral under $P\leftrightarrow Q$, one gets
\ba
0 &=& (d-3)(d-4) \,{\cal S} \;, \nonumber
\ea
which completes the proof of \eq\nr{sunset}.

\subsection{Three-loop bosonic integrals}

\begin{figure}[t]
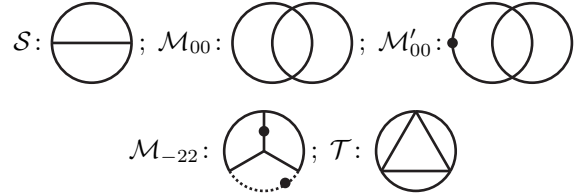

$$ 
{\cal S}\!: \!\ToptVS(\Asc,\Asc,\Lsc) ;\;
{\cal M}_{00}\!: \!\ToprVB(\Asc,\Asc,\Asc,\Asc) ;\;
{\cal M}_{00}^\prime\!: \!\ToprVBdot(\Asc,\Asc,\Asc,\Asc)
$$
$$
{\cal M}_{-22}\!: \ToprVMdot(\Asc,\Ahh,\Asc,\Lsc,\Lsc,\Lsc) ;\;
{\cal T}\!: \TopfVT(\Asc,\Asc,\Asc,\Lsc,\Lsc,\Lsc)
$$
\vspace*{-6mm}
\caption[a]{\it 2-, 3- and 4-loop bosonic master sum-integrals. 
Each solid line corresponds to a bosonic propagator $1/P^2$, each dashed line
to an inverse propagator $P^2$. A dot on a line denotes the same object 
squared.
\label{fig:bosInts}}
\end{figure}

At the 3-loop level, a number of master integrals have been 
encountered in the computation of the pressure. 
For a classification, let us adopt a naming scheme advocated 
in \cite{BN},
\ba
{\cal M}_{ij} &\equiv& 
\sumint_{PQR}\frac1{(P^2)^i\,Q^2\,R^2}
\times\nn&\times&
\frac1{(P-Q)^2\,((Q-R)^2)^j\,(R-P)^2} \;.
\la{Mclass}
\ea
Note that ${\cal M}$ is symmetric in its two indices.
The known cases (cf. \fig\ref{fig:bosInts}) include a 
basketball-type and a spectacles-type integral
\ba
\label{3loopB}
{\cal M}_{00} &\!\!\approx\!\!&
\frac{T^4}{(4\pi)^2}\, \frac1{24\,\e} 
\lk 1+\e\,b_{11}+\e^2\,b_{12}+... \rk \la{b1} \,,\\
{\cal M}_{-22} &\!\!\approx\!\!& 
\frac{T^4}{(4\pi)^2}\, \frac{11}{216\,\e} 
\lk 1+\e\,m_{11}+... \rk \;,
\ea
as well as the modified basketball-type integral that is needed at 
higher orders \cite{phi4},
\ba
{\cal M}_{00}^\prime &\!\!\equiv\!\!& 
\sumint_{PQR}\frac1{(Q^2)^2\,R^2\,(P-Q)^2\,(R-P)^2} \nn
&\!\!\approx\!\!& \frac{T^2}{(4\pi)^4}\, \frac1{8\,\e^2}
\lk 1+\e\,b_{21}+\e^2\,b_{22}+...\rk \,.
\ea
The expansion coefficients read \cite{AZbos,phi4}
\ba
b_{11}&=&\fra{91}{15}-3\gE+8\zA-2\zC \;,\\
b_{21}&=&\fra{17}6+\gE+2\zA \;,\\
b_{22}&=&48.7976.. \;,\\
m_{11}&=&\fra{73}{22}-\fra{21}{11}\gE+\fra{64}{11}\zA-\fra{10}{11}\zC \;,
\ea
while $b_{12}$ has not been computed yet.

\def\TopoVR(#1){\pic{#1(15,15)(15,-90,270)}}
Other integrals of the class \eq\nr{Mclass} can be reduced by standard 
methods, like shifts of integration momenta exploiting the symmetry of the 
integrand, such as
\ba
{\cal M}_{-11} &=& 
-\frac12\,{\cal M}_{00}
+2\,{\cal I}_1^0\,{\cal S} \;, \label{redBos}
\ea
or, in a graphical representation
\ba
\ToprVM(\Asc,\Ahh,\Asc,\Lsc,\Lsc,\Lsc) &\!\!\!=\!\!\!&
-\frac12 \ToprVB(\Asc,\Asc,\Asc,\Asc)
+2 \TopoVR(\Asc)\ToptVS(\Asc,\Asc,\Lsc) \;,\nonumber
\ea
where the notation is like in \fig\ref{fig:bosInts}.

\subsection{Four-loop bosonic integrals}
\la{se:4lbos}

At the 4-loop level, the classification and evaluation of sum-integrals
has not been tackled in a systematic way yet. There is however some 
pioneering work exploring scalar theory at this order \cite{phi4}.
The major simplification (as compared to QCD) is that the reduction step 
is trivial, and there is only one genuine 4-loop sum-integral to compute. 
Hence, the only 4-loop master integral that is known presently is
(cf. \fig\ref{fig:bosInts})
\ba \la{eq:4loopT}
{\cal T}&\!\!\!\equiv\!\!\!&
\sumint_{PQRS}\frac1{P^2(P\!+\!S)^2\,Q^2(Q\!+\!S)^2\,R^2(R\!+\!S)^2} \nn
&\!\!\!\approx\!\!\!&\frac{T^4}{(4\pi)^4}\, \frac1{16\,\e^2} 
\lk 1+\e\,t_{11}+\e^2\,t_{12}+...\rk \;,
\ea
with coefficients
\ba
t_{11}&=&\fra{44}5-4\gE+12\zA
-4\zC-3\zeta(2) \;,\\
t_{12}&=&2b_{12}+25.7055..-3\zeta(2)(28.9250..)\;.
\ea
These coefficients have been calculated in \cite{phi4}, profiting 
from the particular structure of the integral \eq\nr{eq:4loopT}, 
which contains 
three insertions of 1-loop bubble-type integrals, and can hence be 
tackled in complete analogy to the one in \eq\nr{3loopB}, as pioneered
in \cite{AZbos}. The numerical values given above result from simple
numerical integrations and are known to more than 10 digit accuracy.

Note that in the physics calculation where this integral was needed, 
it occurred in combination with the one in \eq\nr{b1} such that the 
number $b_{12}$ cancels in the final result.

\section{Fermionic master integrals}
\la{se:fer}

Let us denote fermionic four-momenta by braces, $\{P\}$, to indicate 
that their $p_0=2\pi T(n+\frac12)$ with $n\in\ZZ$.

There is an important class of relations between fermionic and bosonic
sum-integrals, which can be derived by partitioning the Matsubara sums 
\ba \label{eq:split}
\sum_{n\in\ZZ}=\sum_{n\;{\rm even}}+\sum_{n\;{\rm odd}}
\ea
and then rescaling spatial integration momenta on the left-hand-side 
as $p_i\rightarrow\frac12 p_i$
(see, e.g., \cite{PC}). 

\subsection{One- and two-loop fermionic integrals}

The relations obtained as just described reduce all 
fermionic 1- and 2-loop tadpoles to the respective bosonic case,
\ba \label{eq:ferTad}
\widetilde{{\cal I}}_n^m &\equiv& 
\sumint_{\{P\}} \frac{(p_0)^m}{(P^2)^n} \nn
&=& \(2^{2n-m-3+2\e}-1\) {\cal I}_n^m \;,\\
\la{eq:ferS}
\widetilde{\cal S} &\equiv& 
\sumint_{\{P\}Q} \frac1{P^2\,Q^2\,(P-Q)^2} \nn
&=& \frac13\(2^{4\e}-1\) {\cal S} \;=\; 0 \;,
\ea
where in the last step \eq\nr{sunset} was used. 

\subsection{Three-loop fermionic integrals}

\begin{figure}[t]
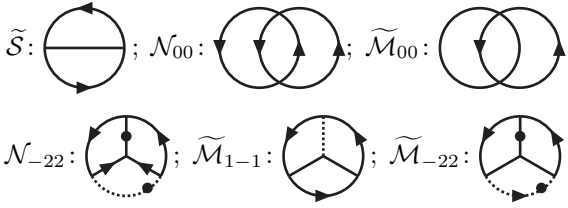

$$
\widetilde{\cal S}\!: \!\ToptVS(\Aqu,\Aqu,\Lsc) ;\;
{\cal N}_{00}\!: \ToprVB(\Aqu,\Aqu,\Aqu,\Aqu) ;\;
\widetilde{\cal M}_{00}\!: \!\ToprVB(\Aqu,\Aqu,\Asc,\Asc)
$$
$$
{\cal N}_{-22}\!: \!\ToprVMdot(\Aqu,\Ahh,\Aqu,\Lqu,\Lqu,\Lsc) ;\;
\widetilde{\cal M}_{1-1}\!: \!\ToprVM(\Aqu,\Aqu,\Aqu,\Lsc,\Lsc,\Lhh)  ;\;
\widetilde{\cal M}_{-22}\!: \!\ToprVMdot(\Aqu,\Agh,\Aqu,\Lsc,\Lsc,\Lsc) 
$$
\vspace*{-6mm}
\caption[a]{\it 2- and 3-loop fermionic master sum-integrals. 
Each arrow-line corresponds to a fermionic propagator.
\label{fig:ferInts}}
\end{figure}

To classify the master integrals at the 3-loop level, 
let us again adopt the notation of \cite{BN},
\ba
{\cal N}_{ij} &\equiv& \sumint_{P\{QR\}}\frac1{(P^2)^i\,Q^2\,R^2}
\times\nn&\times&
\frac1{(P-Q)^2\,((Q-R)^2)^j\,(R-P)^2} \;, \la{Nclass} \\
\widetilde{\cal M}_{ij} &\equiv& \sumint_{\{PQR\}}\frac1{(P^2)^i\,Q^2\,R^2}
\times\nn&\times&
\frac1{(P-Q)^2\,((Q-R)^2)^j\,(R-P)^2} \;.
\ea
Note that, while ${\cal N}$ is still symmetric in its two indices, 
$\widetilde{\cal M}$ is not.

It turns out that, by the same 
partitioning trick as above,
the two possible fermionic 3-loop basketball-type vacuum integrals are
related, such that it suffices to include one of them into the basis of
master integrals,
\ba
\widetilde{\cal M}_{00} &=&
\frac16\(2^{6\e-1}-1\) {\cal M}_{00} - \frac16 {\cal N}_{00} \;, \\
{\cal N}_{00} &\approx&
\frac{T^4}{(4\pi)^2}\,\frac1{96\,\e}\lk 1+\e\,b_{31}+...\rk \;,
\ea
where \cite{AZfer}
\ba
b_{31} &=& \fra{173}{30}-3\gE-\fra{42}5\ln2+8\zA-2\zC \;.
\ea

Further 3-loop masters (see also \fig\ref{fig:ferInts}) are
\ba
{\cal N}_{-22} &\!\!\!\approx\!\!\!& 
\frac{T^4}{(4\pi)^2}\,\frac1{108\,\e}\lk1+\e\,n_{21}+...\rk\;,\\
\widetilde{\cal M}_{1-1} &\!\!\!\approx\!\!\!& 
\frac{T^4}{(4\pi)^2}\(-\frac1{192\,\e}\)\lk1+\e\,m_{21}+...\rk\,,\\
\widetilde{\cal M}_{-22} &\!\!\!\approx\!\!\!& 
\frac{T^4}{(4\pi)^2}\(-\frac{29}{1728\,\e}\)\lk1+\e\,m_{31}+...\rk,
\ea
with \cite{AZfer}
\ba
n_{21} &\!\!=\!\!& \fra{35}{8}-\fra32\gE-\fra{63}{10}\ln2+5\zA-\fra12\zC\;,\\
m_{21} &\!\!=\!\!& \fra{361}{60}+3\gE+\fra{76}5\ln2-4\zA+4\zC\;,\\
m_{31} &\!\!=\!\!& \fra{89}{29}-\fra{39}{29}\gE-\fra{90}{29}\ln2
+\fra{136}{29}\zA-\fra{10}{29}\zC\;.
\ea

In close analogy to the bosonic case \eq\nr{redBos}, 
others get reduced by applying a shift of integration variables and
exploiting the integral's symmetry,
\ba
{\cal N}_{-11} &=& -\frac12{\cal N}_{00}
+2\,\widetilde{\cal I}_1^0\,\widetilde{\cal S} \;,\\
\widetilde{\cal M}_{-11} &=& -\frac12\widetilde{\cal M}_{00}
+{\cal I}_1^0\,\widetilde{\cal S}
+\widetilde{\cal I}_1^0\,{\cal S} \;.
\ea

\subsection{Four-loop fermionic integrals}

\begin{figure}[t]
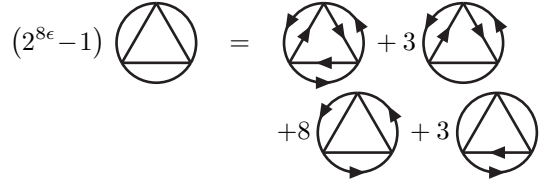

\ba 
\(2^{8\e}\!-\!1\)\TopfVT(\Asc,\Asc,\Asc,\Lsc,\Lsc,\Lsc)
&=&\TopfVT(\Aqu,\Aqu,\Aqu,\Lqu,\Lqu,\Lqu)
+3\TopfVT(\Aqu,\Asc,\Aqu,\Lqu,\Lqu,\Lsc)\nn
&&+8\TopfVT(\Aqu,\Aqu,\Aqu,\Lsc,\Lsc,\Lsc)
+3\TopfVT(\Asc,\Aqu,\Asc,\Lsc,\Lsc,\Lqu)\nonumber
\ea
\vspace*{-6mm}
\caption[a]{\it Graphical representation of a linear relation 
between various 4-loop sum-integrals. Notation is as above.
\label{fig:relT}}
\end{figure}

Not a single genuine fermionic 4-loop sum-integral has been computed to date.
However, at this level, a theoretically important contribution is
expected to be made for the QCD pressure \cite{g6lng}, adding the last 
missing perturbative piece of a physically complete leading-order 
description, which includes (known) non-perturbative coefficients.
Since, naturally, an analytic treatment of diagrams containing quark
loops seems to be a good starting point 
(due to the simpler structure of fermionic vertices and propagators),
progress with this type of sum-integrals can be expected in the 
near future, before the gluonic sector (cf. \se\ref{se:4lbos}) will
be classified.

To obtain a first useful relation, 
one can apply the strategy of \eq\nr{eq:split}
to the bosonic integral of \eq\nr{eq:4loopT}, to get a linear relation 
between the four distinct fermionic integrals of the same topology, as
depicted in \fig\ref{fig:relT}. 

\section{Chemical potential}

In systems which allow for an excess of, say, particles over anti-particles, 
one introduces further parameters -- the {\em chemical potentials} $\mu$ -- 
as Lagrange multipliers of the particle number operators into the Lagrangian.
The effect is to change four-momentum to be used in the free propagators 
of the respective particle species as
\ba
p_0&\rightarrow&p_0-i\mu \;,
\ea
such that all sum-integrals pick up a dependence on the parameter(s) $\mu$.
In the case of QCD, to accommodate for a non-zero baryon number density,
chemical potentials have to be introduced for all flavors of quarks.

In fact, contributions to the QCD pressure have been calculated including 
the effects of a quark chemical potential, requiring a generalization of
the sum-integrals discussed in \se\ref{se:fer} to include the 
$\mu$\/-dependence. This was done through 3-loop order in \cite{AVmu}, 
where all required expansions are shown. While for the 1-loop fermionic 
tadpole an analytic result can simply be given in terms of the generalized 
Zeta function $\zeta(x,z)$ 
(cf. \eq\nr{eq:tadQ} below), the 3-loop generalizations
${\cal N}_{ij}(\mu)$ and $\widetilde{\cal M}_{ij}(\mu)$ are somewhat
more complicated to evaluate, and their expansions turn out to contain
derivatives $\partial_x \zeta(x,z)$ \cite{AVmu}.

One notable structural change as compared to the case $\mu=0$ is that 
certain relations that relied on symmetries of the integrand no longer
hold true. In particular, there are now non-vanishing $\mu$\/-dependent 
2-loop sum-integrals of sunset-type, such as (cf. \eq\nr{eq:ferS})
\ba \la{eq:muS}
\widetilde{\cal S}(\mu) &\approx& -\frac{\mu^2}{(4\pi)^4}\,\frac4\e
\lk 1+\e\,s_{11}(\mu)+...\rk\;,
\ea
with  \cite{AVmu,AGmu}
\ba 
s_{11}(\mu) &\!\!\!=\!\!\!& 2-2\gE+\frac4{1-2z}
[\zeta'(0,z)-\zeta'(0,1-z)]\nn
&\!\!\!=\!\!\!& 2-2\gE+\frac4{2z-1}\ln\frac{\Gamma(1-z)}{\Gamma(z)}\,,
\ea
where $z\equiv\frac12-i\frac{\mu}{2\pi T}$ and 
$\zeta'(x,z)\equiv\partial_x\zeta(x,z)$.
Note that the result \eq\nr{eq:muS} is compatible with \eq\nr{eq:ferS}
in the limit $\mu\rightarrow 0$, as it has to be. 

We will not list the further known integrals here, but finally briefly 
discuss a different setting, in which related sum-integrals appear.

\section{$q$-integrals}

In some cases, one might be interested in integrals which (for $\mu=0$)
interpolate smoothly between the bosonic and fermionic cases, using 
\ba
p_0 &=& 2\pi T(n+q)
\;,\quad n\in\ZZ \;\;\mbox{and}\;\; q\in\RR \;.
\ea
Below, four-momenta with such $q$\/-dependent components will 
be denoted by brackets, $[P]$.
Hence, all sum-integrals pick up a dependence on the parameter $q$,
and, since the integers $n$ are summed, are periodic with respect 
to a unit change in $q$, such that it is sufficient to determine 
them on the interval $q\in[0,1)$. Note that $q=0$ corresponds to 
the bosonic integrals discussed in \se\ref{se:bos}, while $q=\frac12$
corresponds to the fermionic ones discussed in \se\ref{se:fer}.

Notably, the need for such $q$\/-dependent sum-integrals arises
in attempts to analyze the behavior of the so-called spatial 
't Hooft loop \cite{thooft},
which can be taken as an order parameter for the deconfinement 
phase transition in pure Yang-Mills gauge theory, and requires 
the minimization of a certain effective action with respect to the
parameter $q$ \cite{chris}.

To show a concrete example, the basic bosonic 1-loop tadpole 
can now be represented as
\ba \la{eq:tadQ}
{\cal I}_n^m(q) &\equiv& \sumint_{[P]} \frac{(p_0)^m}{(P^2)^n} \nn
&=& \frac{\pi^2 T^4}{(2\pi T)^{2n-m}}
\frac{4^\e\,\Gamma(n-\frac32+\e)}{\Gamma(\frac12)\Gamma(n)}\,
\times\nn&\times&
[(-1)^m\zeta(2n-m-3+2\e,1-q)+
\nn&&{}
 +\zeta(2n-m-3+2\e,q)]\;,
\ea
where $\zeta(x,q)=\sum_{n=0}^\infty\frac1{(n+q)^x}$ is the generalized 
Zeta function, with $\zeta(x,1)=\zeta(x)$ and 
$\zeta(x,\frac12)=(2^x-1)\zeta(x)$ relating ${\cal I}_n^m(q)$ to
\eq\nr{eq:bosTad} and \eq\nr{eq:ferTad}.

We will not discuss higher loop orders here, although some results are 
also available in the literature \cite{chris}, while others (notably up to
the 3-loop level) can in principle be deduced from \cite{AVmu,AGmu} 
by analytically continuing the $q$\/-dependence
to complex values $q\rightarrow z=\frac12-i\mu/(2\pi T)$, such as
the $q$\/-dependent 2-loop sunset-type sum-integral
\ba
{\cal S}(q) &=& \frac{T^2}{(4\pi)^2}\,\frac{(2q-1)^2}{4\,\e} 
\lk 1+\e\,s_{11}(q)+...\rk \;,\nn
s_{11}(q) &=& 2-2\gE+\frac4{2q-1}\ln\frac{\Gamma(1-q)}{\Gamma(q)}\,,\nonumber
\ea
as obtained by explicit calculation, or by analytic continuation from 
\eq\nr{eq:muS}. It vanishes as $q\rightarrow\frac12$, as it should 
(cf. \eq\nr{eq:ferS}).

\end{document}